\documentclass[doublespacing]{narfront}

\newcommand{\DG}{$\Delta G$ }

\newcommand{\commentout}[1]{}

\usepackage{graphics}

\usepackage{amssymb}
\usepackage{multirow}
\usepackage{slashbox}
\usepackage{rotating}

\newcommand{\html}[1]{{ }} 
\newcommand{\htmladdnormallink}[1]{{ }} 

\usepackage{color,soul}

\begin{document}
\sethlcolor{white}  

\begin{frontmatter}

\title{The effects of mismatches on hybridization in DNA microarrays:
determination of nearest neighbor parameters}

\author[label1,label2,cor]{J. Hooyberghs},
\ead{jef.hooyberghs@vito.be}
\author[label3]{P. Van Hummelen},
\ead{paul.vanhummelen@vib.be}
\and
\author[label1]{E. Carlon},
\corauth[cor]{Corresponding author.}
\ead{enrico.carlon@fys.kuleuven.be}

\address[label1]{Institute for Theoretical Physics, Katholieke
Universiteit Leuven, Celestijnenlaan 200D, B-3001 Leuven, Belgium}

\address[label2]{Flemish Institute for Technological Research (VITO), Boeretang 200, B-2400 Mol, Belgium}

\address[label3]{MicroArray Facility, VIB, Herestraat 49, B-3000 Leuven, Belgium}

\begin{abstract}
\textbf{ 
Quantifying interactions in DNA microarrays is of central importance
for a better understanding of their functioning.  Hybridization
thermodynamics for nucleic acid strands in aqueous solution can be
described by the so-called nearest-neighbor model, which estimates
the hybridization free energy of a given sequence as a sum of
dinucleotide terms. Compared with its solution counterparts,
hybridization in DNA microarrays may be hindered due to the presence
of a solid surface and of a high density of DNA strands. We present
here a study aimed at the determination of hybridization free
energies in DNA microarrays. Experiments are performed on custom
Agilent slides. The solution contains a single oligonucleotide. The
microarray contains spots with a perfect matching complementary
sequence and other spots with one or two mismatches: in total 1006
different probe spots, each replicated 15 times per microarray. The
free energy parameters are directly fitted from microarray data. The
experiments demonstrate a clear correlation between hybridization
free energies in the microarray and in solution.  The experiments
are fully consistent with the Langmuir model at low intensities, but
show a clear deviation at intermediate (non-saturating) intensities.
These results provide new interesting insights for the
quantification of molecular interactions in DNA microarrays.
}
\end{abstract}


\end{frontmatter}

\bibliographystyle{nar}

\section*{Introduction}
\label{sec:intro}

DNA microarrays are
{ widely} used in the current research in molecular biology
\cite{brow99}. Such devices have several important applications
\cite{stou05} as for instance in gene expression profiling, in the
detection of Single Nucleotide Polymorphisms, in the analysis of
copy number variations and of target sequences for transcription
factors. Several different platforms, either commercial or home
made, are currently available. They differ by the details of
fabrications (via spotting or in situ growth), the length of the
sequences (oligonucleotides or long PCR fragments) and the chemistry
of fixation.  What all DNA microarrays have in common is the basic
underlying reaction of hybridization between a nucleic acid strand
in solution and a complementary strand linked covalently at a solid
surface.  Hybridization is characterized by a (sequence dependent)
free energy difference \DG which measures the binding affinity for
the two strands to form a duplex.

In the past decades a large number of papers were dedicated to the
investigation of static and dynamic properties of the hybridization
between nucleic acid strands which are both floating in an aqueous
solution (see \cite{bloo00} and references therein). Nearest neighbor
models provide resonable approximation of \DG for strands hybridizing
in solution \cite{bore74,frei86}. In these models \DG is
{ calculated} as a sum of ``stacking" parameters associated to
dinucleotides \cite{bloo00}. The nearest neighbor model is known to
be rather accurate at least for hybridization between complementary
strands. The case of single internal mismatches \cite{sant04} as
well as the dependence of \DG on other parameters as the monovalent
salt concentration \cite{sant98} were also considered.

There has been some discussion in the literature about the
relationship between hybridization in solution and hybridization in
DNA microarrays. In early studies of gel pad microarrays
\cite{foti98} a linear relationship between microarray hybridization
free energies ($\Delta G_{\rm \mu array}$) and the corresponding
free energies in solution ($\Delta G_{\rm sol}$) was found. Recently
\cite{weck07} a similar relationship was observed on self-spotted
codelink activated slides. Other studies on Affymetrix Genechips
\cite{naef03,zhan03} report very weak correlation between $\Delta
G_{\rm \mu array}$ and $\Delta G_{\rm sol}$. In some papers
\cite{held03,carl06} however the same Affymetrix data could be
fitted resonably well with a linearly rescaled $\Delta G_{\rm sol}$.
Also some recent measurements of thermodynamic parameters using a
temperature dependent surface plasmon resonance \cite{fich07} seem
to suggest a decreased $\Delta G_{\rm \mu array}$ compared to
$\Delta G_{\rm sol}$. Clearly, as also some recent literature points
out \cite{sant04,fish07,pozh07}, more systematic physico-chemical
studies are required for a better understanding of hybridization in
DNA microarrays.  A precise quantification of \DG is important.
Through a better understanding of molecular interactions between
hybridizing strands it would be possible to turn microarrays into
more precise tools for large scale genomic analysis. For instance
one could estimate gene expression levels or detect mutations
through an analysis based on thermodynamics instead of using
empirical statistical methods.

This paper is dedicated to the investigation of the applicability
of the nearest neighbor model to describe hybridization
{ reactions} in DNA microarrays{, with a focus on sequences
that contain isolated mismatches}. Experimental results involving
the hybridization of one sequence in solution with a large set of
different sequences on a microarray will be presented. The stacking
free energy parameters will be determined from the analysis of the
behavior of the experimental fluorescent intensities measured from
different spots of the microarray.
We will be interested in the correlation between { free energies
resulting from} these parameters and the equivalent { quantities
calculated from experimental stacking free energy parameters of
nucleic acid melting in aqueous solution}.
The analysis of the experimental data clearly reveals a good degree
of correlation.
{ However}, a much better agreement with thermodynamic models is
found if the thermodynamic parameters are directly fitted from the
experimental microarray data. In addition { to this tight
agreement with theory}, a regime is found where the data are cleary
deviating from the Langmuir behavior.

This paper is organized as follows. Materials and Methods discusses
the experimental setup, the thermodynamic model of hybridization and
the fitting procedure. In Results and Discussion the experimental
results are presented and a comparison between free energies fitted
from the microarray data and their solution counterparts is done.
The final part of the paper is dedicated to a general discussion in
which some open issues are highlighted.

\begin{table}[t]
\caption{The oligos used as target in the four different
hybridization expriments. The oligos were bought from Eurogentec in
duplicate obtained from independent synthesis cycles.}
\begin{tabular}{lll}
  \hline
  Name & Sequence & Labeling \\
  \hline
  Target1 & 5' GTTTTCGAAGATTGGGTGGCACTGTTGTAA 3' & 20-mer poly A + Cy3  on 3' \\
  Target2 & 5' CAGGGCCTCGTTATCAATGGAGTAGGTTTC 3' & 20-mer poly A + Cy3  on 3' \\
  Target3 & 5' CTTTGTCGAGCTGGTATTTGGAGAACACGT 3' & 20-mer poly A + Cy3  on 3' \\
  Target4 & 5' GCTTCTCCTTAATGTCACGCACGATTTCCC 3' & 20-mer poly A + Cy3  on 3' \\
  \hline
  \label{table_targetseq}
\end{tabular}
\end{table}

\section*{Materials and Methods}

\subsection*{The design of the experiment}

%

For the present study several hybridization experiments were
performed, each with a single oligonucleotide sequence (referred to
as the target in this paper) in solution at different
concentrations. Four different targets were used in the experiments,
and their sequences are given in Table \ref{table_targetseq}. The
sequences contain a 30-mer hybridizing stretch followed by a 20-mer
poly(A) spacer and a Cy3 label at the 3' end of the sequence. Each
target oligo was bought in duplicate, in order to check the quality
of the target synthesis. In the rest of the paper we will refer to
the two duplicated oligos as a and b.

The sequences printed at the microarray surfaces and referred
{ to}
here as
the probes, were chosen to contain up to two mismatches, following the
scheme shown in Table \ref{table_probeset}. Mismatches were inserted
from nucleotides 6 to 25 along the 30-mer sequences in order to
avoid terminal regions. In the probes with two mismatches these were
separated by at least 5 nucleotides. Given the nucleotide of the target
strand there are three different possible mismatching nucleotides and
20 available positions, hence in total 60 single mismatch sequences. A
similar counting for double mismatches yields 945 different sequences
(see Table \ref{table_probeset}). The total number of probe sequences,
including the perfect matching one, is 1006.

\begin{table}[t]
\caption{Design of probeset: probe sequences covalently linked at the
microarray surface contained up to two mismatches following the scheme
shown in this table. In total there are 1006 different probe sequences,
replicated 15 times in the custom 8$\times$15K custom Agilent slide.}
\begin{tabular}{lll}
  \hline
  Nr of probes & type of mismatch & location of mismatch \\
  \hline
  1 & perfect match & --- \\
  60 & single mismatch (all 3 permutations) & site 6 to 25 \\
  945 & double mismatch (all 9 permutations) & site 6 to 25, separated by min. 5 sites\\
  \hline
\end{tabular}
  \label{table_probeset}
\end{table}

For each experiment one target and one 8x15K custom Agilent slide was
used. This slide consists of eight identical microarrays and each of
these can contain up to more than 15 thousand spots. The 1006 probe
sequences were spotted in the custom array 15 times: in 12 repicates
a 30-mer poly(A) was added on the 3' side (surface side), in order
to asses the effect of a sequence spacer. Three replicates contained
no poly(A) spacer.  The
{ eight} microarrays of one slide have to be hybridized during
the same experiment, but a different target solution can be used. In
the experiments the target concentrations ranged from $50$ to
$10,000$ pM (picomolar) according to the scheme given in Table
\ref{table_targetconc}. In experiment 1 only target a was used,
while in the experiments 2, 3 and 4 both replicated targets (a and
b) were used. Finally, in experiment 1 and 2 a fragmentation of the
target was performed before hybridization (see section on
hybridization protocol for details).

\begin{table}
\caption{The target condition per microarray: concentration, oligo
synthesis a or b, fragmentation f if applied .}
\begin{tabular}{cllll}
  \hline
 Microarray  & Experiment/target 1 & Experiment/target 2 & Experiment/target 3 & Experiment/target 4 \\
  \hline
  1 & 10000 pM, a, f & 10000 pM, a, f & 10000 pM, a & 1000 pM, a \\
  2 & 7500 pM, a, f& 5000 pM, a, f & 5000 pM, a & 500 pM,a  \\
  3 & 5000 pM, a, f & 1000 pM, a, f & 1000 pM, a & 100 pM, a \\
  4 & 2500 pM, a, f & 50 pM, a, f & 50 pM, a & 50 pM, a  \\
  5 & 1000 pM, a, f & 10000 pM, b, f & 10000 pM, b & 1000 pM, b \\
  6 & 500 pM, a, f & 5000 pM, b, f & 5000 pM, b & 500 pM, b \\
  7 & 100 pM, a, f & 1000 pM, b, f & 1000 pM, b & 100 pM, b \\
  8 & 50 pM, a, f & 50 pM, b, f & 50 pM, b & 50 pM, b \\
  \hline
  \label{table_targetconc}
\end{tabular}
\end{table}

The four 30-mer target sequences were selected from fragments of human
genes having a GC content ranging from 43\% to 50\%. A criterion for
selecting the target sequences was the requirement that the probes
constructed following the scheme in Table \ref{table_probeset} would
yield a roughly flat histogram of mismatch types, so that all mismatches
are approximately equally present in the experiments.

\subsection*{Hybridization protocol and scanning}

{
For the experiments
we
used the commercially available Agilent platform and followed a
standard protocol with Agilent products, as described below. (The target
oligonucleotides were OliGold$\circledR$ from Eurogentec, Seraing,
Belgium). Hybridization mixtures contained one target
oligonucleotide with a 3' Cy3 endlabeling diluted in nuclease-free
water to the final concentration together with 5 $\mu$L 10x blocking
agent and 25 $\mu$L 2x GEx hybridization buffer HI-RPM.
Unfortunately Agilent Techologies does not disclose the precise
composition of the hybridization buffer in the content of salt
and other chemicals.}
In experiment 1 and 2 the addition of the hybridization buffer was
proceeded by a fragmentation step, 1 $\mu$L fragmentation buffer was
added followed by an incubation of 30 min at 60$^\circ$C.  { This
fragmentation buffer is customarily used in Agilent hybridization
platforms and produces targets sequences of reduced length in order
to speed up the hybridization reaction. Too long sequences, as
obtained from biological extracts, e.g. from reverse transcription
of mRNA samples, have a reduced hybridization efficiency due to
steric hindrance. By comparing experiments with and without
fragmentation we found that the fragmentation step has little effect
on the results. (More information can be found in the online
supplementary material.)} The hybridization mixture was centrifuged
at 13000
{ rpm} for 1 min and each microarray of the 8x15K custom Agilent
slides was loaded with 40 $\mu$L. The hybridization occurred in an
Agilent oven at 65$^\circ$C for 17 hours with rotor setting 10 and
the washing was performed according to the manufacturer's
instructions. The arrays were scanned on an Agilent scanner
(G2565BA) at 5 $\mu$m resolution, high+low laser intensity and
further processed using Agilent Feature Extraction Software (GE1 v5
95 Feb07) which performs automatic gridding, intensity measurement,
background subtraction and quality checks.

\subsection*{Thermodynamic Model}

In the Langmuir model the dynamics of hybridization is described by a
rate equation for $\theta$, the fraction of hybridized probes from a
spot as follows

\begin{equation}
\frac{d \theta}{dt} = c k_1 (1-\theta ) - k_{-1} \theta
\end{equation}
where $c$ is the target concentration and $k_1$ and $k_{-1}$ are the
attachment and detachment rates. The equilibrium value for $\theta$
can be obtained from the condition $d\theta_{eq}/dt=0$. Using the link
between the rates and equilibrium constants, i.e. ${k_1}/{k_{-1}} =
e^{-\Delta G/RT}$, with \DG the hybridization free energy, $R$ the gas
constant and $T$ the temperature one finds

\begin{equation}
\theta_{eq} = \frac{ c \ e^{-\Delta G /RT}}{1 + c \ e^{-\Delta G /RT}}
\label{langmuir}
\end{equation}
which is the so-called Langmuir isotherm. To link this isotherm to the
measured quantities one assumes that the fraction of hybridized probes
is linearly related to the measured fluorescent intensity measured from
a spot, which yields

\begin{equation}
I = \frac{ A c \ e^{-\Delta G /RT}}{1 + c \ e^{-\Delta G /RT}}
\label{langmuir_I}
\end{equation}
{ Here $I$ is the background-subtracted intensity, where the
background subtraction, as explained above is done by Agilent
Feature Extraction software. In the rest of the paper we will no
longer explicitly state that the intensities are background
subtracted, and will simply refer to them as intensities.}
$A$ is a constant which is an overall scale factor. Far from
chemical saturation, i.e. when only a small fraction of surface
sequences is hybridized (i.e. $c \ e^{-\Delta G /RT} \ll 1$) one can
neglect the denominator in Eq.~(\ref{langmuir}) to get:

\begin{equation}
I \approx A c \ e^{-\Delta G /RT} \label{IvsDG}
\label{langmuir_linear}
\end{equation}

In the nearest neighbor model the hybridization free energy of
perfect complementary strands is approximated
as a sum of
dinucleotide terms. For instance:

\begin{equation}
\Delta G \left({{\rm ATCCT} } \atop {{\rm TAGGA} } \right) = \Delta
G \left({\rm AT} \atop {\rm TA}  \right) + \Delta G \left({\rm TC}
\atop {\rm AG}  \right) + \Delta G \left({\rm CC} \atop {\rm GG}
\right) + \Delta G \left({\rm CT} \atop {\rm GA} \right) + \Delta G
_{\rm init} \label{DGpm}
\end{equation}
where $\Delta G _{\rm init}$ is an initiation parameter. Since we will
only consider differences of $\Delta G$ between a perfect matching
hybridization and
{ a} hybridization with one or multiple mismatches (see
Eq.~(\ref{how_to_fit})), this initiation parameter will not
contribute and it is omitted in the rest of the paper.  For DNA/DNA
hybrids, symmetries reduce the number of independent parameters to
10 \cite{bloo00}. The nearest neighbor model can be extended to
include single internal mismatches; as an example we consider the
free energy of a stretch with an internal mismatch of CT type

\begin{equation}
\Delta G
\left({{\rm AT{\underline{C}}CT} } \atop {{\rm TA{\underline{T}}GA} }
\right) =
\Delta G \left({\rm AT} \atop {\rm TA}  \right) +
\Delta G \left({{\rm T{\underline C}}} \atop {{\rm A{\underline T}}}  \right) +
\Delta G \left({{\rm G{\underline T}}} \atop {{\rm C{\underline C}}}  \right) +
\Delta G \left({\rm CT} \atop {\rm GA} \right)
\label{DGmm1}
\end{equation}
{ The mismatching nucleotides are underlined and for notational
reasons the mismatch is always put in the second part of the
dinucleotide (which requires the use of symmetry like here in
dinucleotide term three) }. There are 12 types of mismatches and 4
types of flanking nucleotide pairs, hence in total there are 48
mismatch parameters of dinucleotide type.

%

{ There are several possible ways of extracting the $48+10$
dinucleotide parameters from the experimental data. One can either
fit the full Langmuir isotherm} (Eq.~(\ref{langmuir})) {, or for
experiments at sufficiently low concentrations one could consider
the limiting case of} Eq.~(\ref{langmuir_linear}) {. In addition,
the parameters could be extracted either from an experiment at fixed
concentration $c$, by comparing the intensities of different probe
sequences, or from experiments at different concentrations by
analyzing the intensities of identical probe sequences over a
concentration range. As argued in the next sections the optimal
strategy in our experimental setup is to fit}
Eq.~(\ref{langmuir_linear}) {at a fixed low concentration (the
supplementary online material discusses other strategies).}

Equation~(\ref{IvsDG}) contains the constant $A$ which is an overall
scale factor relating the hybridization probability to the actual
measured fluorescence intensity. This quantity may fluctuate from
experiment to experiment. { For instance, the optical scanning
influences $A$, as this is proportional to the laser intensity used.
Also hybridizations in different slides might occur at slightly
varying conditions and there can be small differences in the
manifacturing of the slides. } { In the rest of this paper we
will focus on relative intensities and relative free energies, i.e.
for each microarray we will use the perfect match of that microarray
as a point of reference. We denote the logarithmic ratios of the
intensities with the perfect match intensity as}

\begin{equation}
y_i=\ln I_i-\ln I_{PM}=-\frac{\Delta G - \Delta G_{PM}}{RT} \equiv
-\frac{\Delta \Delta G}{RT} \label{how_to_fit}
\end{equation}

for which the exact value of $A$ is irrelevant and we only need to
consider the relative free energy differences $\Delta \Delta G$
(which is for each probe a positive number).
In $\Delta \Delta G$ of a duplex, only dinucleotide
parameters which are flanking a mismatch remain, the other
parameters cancel out in the subtraction. E.g. from Eq.~(\ref{DGpm})
and (\ref{DGmm1}) one gets

\begin{equation}
\Delta \Delta G \left({{\rm AT{\underline{C}}CT} } \atop {{\rm
TA{\underline{T}}GA} } \right) = \Delta G \left({{\rm T{\underline
C}}} \atop {{\rm A{\underline T}}}  \right) + \Delta G \left({{\rm
G{\underline T}}} \atop {{\rm C{\underline C}}}  \right) - \Delta G
\left({\rm TA} \atop {\rm AT}  \right) - \Delta G \left({\rm AC}
\atop {\rm TG}  \right) \label{deltadelta}
\end{equation}

In this equation the { lower} strand refers to the target
sequence in solution{, which is fixed}. The { upper} strand is
that of the probe sequence attached to the solid surface. Hence, the
$\Delta \Delta G$ of a { duplex} contains two mismatch dinucleotide
parameters and two matching dinucleotide parameters per mismatch. This
holds for sequences that contain more mismatches as long as the nearest
neighbor model is valid, which we assume in our setup since mismatches are
separated at least by five base pairs.  The model can now be written as

\begin{equation}
y_i = \sum_{\alpha=1}^{58} X_{i \alpha} \frac{\Delta G_{\alpha}}{RT}
\end{equation}

where $\alpha$ is the index running over the 58 possible dinucleotide
parameters and $X$ is a frequency matrix, whose elements $X_{i \alpha}$
are the number of times the dinucleotide parameter $\alpha$ enters in
$\Delta \Delta G$ of probe sequence $i$.  With a simple extension of
matrices and vectors one can rewrite the problem as

\begin{equation}
\vec{y} = X \vec{\omega} \label{lin_fit}
\end{equation}
where we have defined $\omega_\alpha = \Delta G_\alpha/RT$.
Having written the problem in Eq.~(\ref{lin_fit}) as a linear one,
we can now apply the standard approach to find the optimal values of
the parameters. The procedure consists in minimizing $S = (\vec{y} -
X \vec{\omega})^2$, which amounts to solving the following linear equation

\begin{equation}
X^T (\vec{y}-X\vec{\omega})=0
\label{lin_opt}
\end{equation}
where $X^T$ is the
{ transpose} of $X$.

\subsection*{Degeneracies of $\vec \omega$}

To obtain $\vec{\omega}$ from
Eq.~(\ref{lin_opt}) one has to invert the $58 \times 58$ matrix $X^T
X$. In the case that $X^T X$ is not invertible one applies a
singular value decomposition \cite{gray97a}.  In the present case
the matrix is not invertible. Zero eigenvalues of the matrix $X^T X$
come from reparametrizations that leave the physically accessible
parameters $\Delta \Delta G$ invariant.
It is known, indeed, that the dinucleotide mismatch parameters are
not uniquely determined \cite{peyr99,gray97a}, as these parameters
are entering in the expression for the total \DG in pairs (see
Eq.~(\ref{DGmm1})). For instance, a reparametrization of the type:

\begin{equation}
\Delta G' \left({{\rm x \,\, {\underline C}}}
\atop {{\rm x'\, {\underline T}}}  \right) =
\Delta G \left({{\rm x \,\, {\underline C}}} \atop
{{\rm x'\, {\underline T}}}  \right) + \varepsilon
\hspace*{12mm} \Delta G' \left({{\rm y \,\, {\underline
T}}} \atop {{\rm y'\, {\underline C}}}  \right) = \Delta G \left({{\rm
y \,\, {\underline T}}} \atop {{\rm y'\, {\underline C}}}  \right)
- \varepsilon
\label{reparam}
\end{equation}
for every pair of complementary nucleotides {$x,x'$ and $y,y'$} leaves the
total \DG invariant, as it can be verified directly from
Eq.~(\ref{DGmm1}). Similar reparametrizations are possible for
mismatches of type AG, AC and TG. Next to these there are three more
invariances which involve a reparametrization of both mismatch and
matching dinucleotide parameters. Hence one has at least $7$ zero
eigenvalues in $X^T X$. { A more detailed discussion of
degeneracies of $X^T X$ can be found in the supplementary online
material. }

\section*{Results and Discussion}

\subsection*{Control of the quality of the experiments}

\begin{figure}[ht]
\begin{center}
\rotatebox{0}{\scalebox{0.80}{\includegraphics*{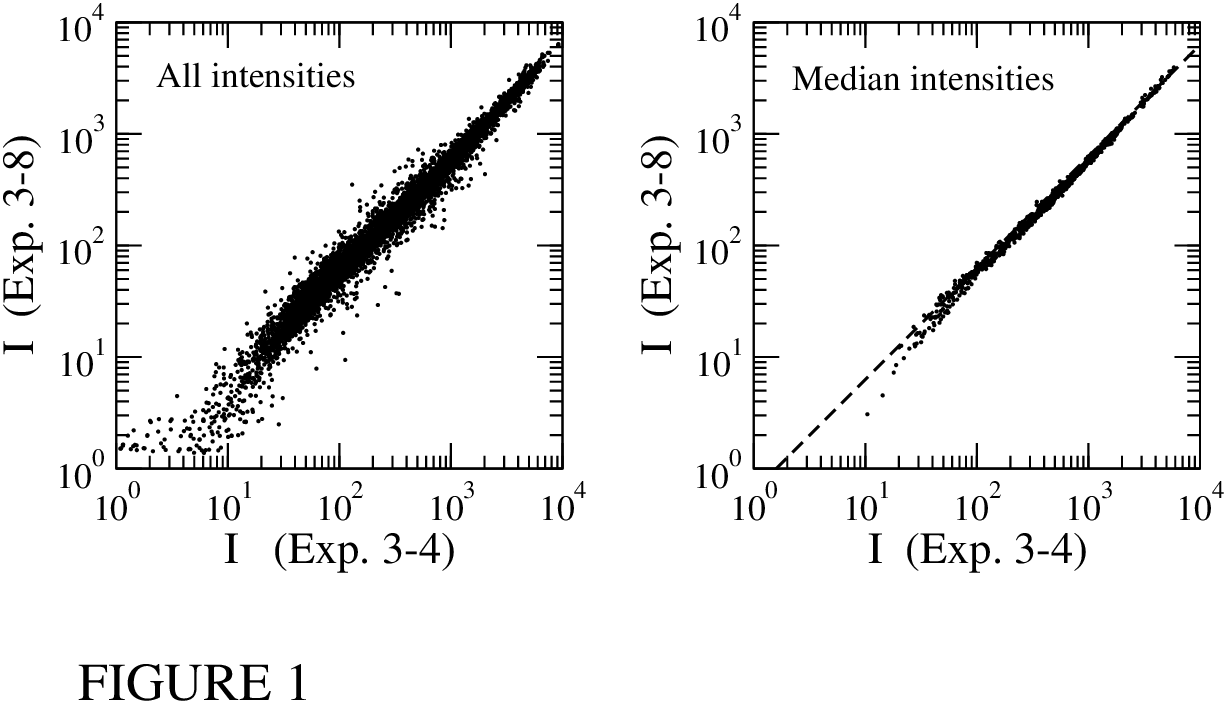}}}
\end{center}
\caption{Correlation plots for intensities in two replicated experiments
at 50 pM for oligo 3a (x-axis) and oligo 3b (y-axis); these are the
experiments 3-4 and 3-8 in Table \ref{table_probeset}. The left plot
shows the total intensities and the right plot concerns only the median
intensities taken for the 15 replicated spots. The dashed line has slope
equal to one.}
\label{fig01}
\end{figure}

As a control of the reproducibility of the result we consider the
intensities correlation between analogous spots in replicated
experiments. The replicated hybridizations were carried out on two
microarrays of the same slide, with two identical but separately
synthesized and labeled target oligos, at the same manually prepared
concentration in solution, see Table \ref{table_targetconc}.
Figure~\ref{fig01} is an example thereof. It shows correlation plots
between two replicated hybridizations. Two plots are shown, one with
the full 15K intensities (left) and one in which the median of the
intensities of the 15 replicated spots are taken (right). In the
former some data spreading is observed, which is greatly reduced
when the median over 15 replicated spots is taken. Note that the
experimental data do not align perfectly on the diagonal of the
graph, this may be attributed to the manual preparation of the
solutions or to differences in the oligos (synthesis or labeling).
Data from different microarrays are aligned on a line of slope equal
to one in the log-log plots of Fig.~\ref{fig01}, which implies a
linear relationship between the intensities. In general, replicates
show a strong correlation between median intensities, which is an
indication of a good reproducibility of the results. { We
included in this median the probes with and without poly(A) spacer.
No significant difference was found in the intensities from spots
with poly(A) and without poly(A) spacer. From this point on, the
median intensity of 15 replicates is always used and simply referred
to as the intensity of a probe, and because of the good
reproducibility we will only discuss the data produced by
hybridizations with oligo synthesis a (see Table }
\ref{table_targetconc}) .

\subsection*{Data analysis with $\Delta G_{\rm sol}$}

{ Next, we consider the relation between the intensities and the
corresponding $\Delta G_{\rm sol}$ for hybridizations in solution
with one or two mismatches.} In the case of two mismatches $\Delta
G_{\rm sol}$ was calculated as the sum of nearest neighbor
parameters for individual mismatches, assuming that the presence of
two mismatches does not involve additional terms in the free energy,
i.e. they do not interact. {In the experiment}
the minimal distance between two mismatches is 5 nucleotides,
which is considered sufficient, in first approximation to support
the non-interaction assumption. In the calculation of \DG from the
tabulated values of $\Delta H$ and $\Delta S$ the temperature was set
to the experimental value  $T=65 ^\circ C$.

Figure~\ref{fig02}{ (a)} shows plots of the intensities vs.
$\Delta \Delta G_{\rm sol}$ as taken from the nearest neighbor model
with the existing tabulated values for hybridization in solution
(see Ref.~\cite{sant04} and references therein). $\Delta \Delta
G_{\rm sol}$ is obtained by subtracting from all free energies that
of the PM sequence, which is taken as a reference.
{As a consequence, for the PM intensities $\Delta \Delta G_{\rm
sol}= 0$. } Each plot in Fig.~\ref{fig02} contains 1006 data points
obtained from the median value of the 15 replicated spots on each
array.

{ As it is well-known from several studies of
melting/hybridization in aqueous solution (see e.g.} \cite{sant98}),
{ the hybridization free energy $\Delta G_{\rm sol}$ depends on
the buffer conditions, and in particular of the ionic strenght of
the solution. Particularly studied was the effect of salt
concentration (NaCl), which is usually assumed to be independent of
sequence, but to be dependent on oligonucleotide length. Melting
experiments in solution are consistent with the following dependence
on Na ions concentrations} \cite{sant98}
\begin{equation}
\Delta G_{\rm sol} = \Delta G_{\rm sol}
(1 M [Na^+]) - a N \ln [Na^+]
\label{DG_salt}
\end{equation}
{ where $\Delta G_{\rm sol} (1 M [Na^+])$ is measured at 1M NaCl,
$N$ is the number of phosphates in the sequence and $a$ a constant.
To our knowledge, the salt effect on sequences with internal
mismatches has not been investigated yet, as measurements were done
at 1M NaCl } \cite{sant04}. {However, salt has mostly an effect
on interactions with the negatively charged phosphate molecules. It
is hence plausible to expect the same type of correction as}
Eq.~(\ref{DG_salt}) {also for sequences carrying mismatches. If
that is the case, the salt dependence cancels out from $\Delta
\Delta G_{\rm sol}$, which is the quantity we are interested in. In
the rest of the paper, we will set the value at 1M NaCl in $\Delta
G_{\rm sol}$.}


{ Figure} ~\ref{fig02}(a) {shows the data for Experiment 1 at
three different concentrations, from bottom to top of $50$, $500$
and $5000$ pM. When plotted as functions of $\Delta \Delta G_{\rm
sol}$ the data points tend to cluster along single monotonic curves.
This already suggests a fair degree of correlation between $\Delta
G_{\rm sol}$ and $\Delta G_{\rm \mu array}$. The experiment at
$5000$ pM shows a pronounced saturation of the intensities, as
expected from the Langmuir model } (Eq.~\ref{langmuir}). {
Sufficiently far from saturation one expects a linear relationship
between the logarithm of the intensity and $\Delta G$, as given by}
Eq.~(\ref{langmuir_linear}). Figure~\ref{fig02} { shows that the
low concentration data at low intensities follow approximately a
straight with the slope $1/RT$ expected from equilibrium
thermodynamics at $T=65 ^\circ C$, which is the experimental
temperature.}

{However, the global behavior of the three concentrations is at
odds with the Langmuir model, which predicts that Intensity vs. free
energy plots for different concentrations should saturate at a
common intensity value $A$, as indicated in} Fig.~\ref{fig02}(b).
{ Although one may expect some variations on $A$ from experiment
to experiment, the data of} Figure~\ref{fig02}(a) {are hard to
reconcile with the Langmuir model. We conclude that the
hybridization data deviate from the full Langmuir model of}
Eq.~(\ref{langmuir}), {but they are in rather good agreement with
its limiting low intensities behavior} Eq.~(\ref{langmuir_linear}).
{ In order to obtain estimates of the free energies $\Delta
\Delta G_{\rm \mu arrays}$ from microarray data we will use then}
Eq.~(\ref{langmuir_linear}) {and restrict ourselves to the lower
concentration data. The analysis of the higher concentration regime
is presented in the supplementary online material.}


\subsection*{Fitting the free energy parameters}

{To fit the 58 parameters of the nearest neighbor model we use
the lowest concentration data, i.e. 50 pM.  Hereto we applied the
algebraic procedure explained in Materials and Methods, which fits
the logarithm of the ratios $I/I_{PM}$ and which assumes that the
data can be described by} Eq.~(\ref{langmuir_linear}). {For low
concentrations this assumption is expected to be correct for the
lower intensities but not for the highest intensities, which deviate
from the Langmuir isotherm as shown in} Fig.~\ref{fig02}.
{This poses a problem for the
fitting procedure since it was designed with the perfect match
intensity $I_{PM}$ as a reference} (Eq.~(\ref{how_to_fit})).
{One may think to circumvent this problem by restricing the fit
to low intensities, for instance only to probes with two mismatches
and rewrite} Eq.~(\ref{how_to_fit}) {using as reference not
$I_{PM}$, but for instance one of the intensities of a probe with
two internal mismatches. This procedure turns out to be of little
practical use for our purposes which is to estimate the free energy
difference between perfect matching sequences and sequences with one
or multiple mismatches and for which the PM reference value is necessary
(a more detailed discussion is in the online material).}

{From the analysis of plots of Intensity vs. $\Delta \Delta G_{\rm
sol}$} (Figure~\ref{fig02}) {one finds that the PM intensity is
systematically lower than that predicted by} (Eq.~(\ref{langmuir_linear})),
{which is the straight line in} Figure~\ref{fig02}(a).
{Hence, the relative intensities $I/I_{PM}$ of the probes that
contain mismatches
are systematically higher than those predicted by}
Eq.~(\ref{langmuir_linear}). {Consequently, a direct fit of the
experimental data to} Eq.~(\ref{how_to_fit}) {underestimates the
effect of a mismatch, which will result in free energy penalties
that are too small. The result of the fit is shown in Figure}
\ref{fig03}. {One can notice that the $\Delta \Delta G$ range is
indeed smaller than the one from hybridization in solution} (Figure
\ref{fig02}). {Moreover, the underestimation of $\Delta \Delta G$
is more severe for probes with two mismatches than for those with
only one, since $\Delta \Delta G$ is a sum of contributions per
mismatch. This produces a discontinuity of the curve from double to
single mismatches. The appearance of this discontinuity is another
evidence of the fact that} Eq.~(\ref{langmuir_linear}) {is not
valid in the full range of intensities. }

{In order to solve this problem, one would need to fit the
data with a more general model $I(c,\Delta G)$ that incorporates
the observed deviations from} Eq.~(\ref{langmuir_linear}). {
As mentioned above, and as shown explicitely from the data analysis
in the supplementary online material, the deviations cannot be
described within the general Langmuir model} (Eq.~(\ref{langmuir})).
%
{At present, it is not yet clear which alternative model to use
for $I(c,\Delta G)$. Moreover, the choice of this model may
considerably influence the fitted nearest neighbor parameters. A
safer compromize is to start from the observation that}
Eq.~(\ref{langmuir_linear}) {is followed by the large majority of
the low concentration data points in} Fig.~\ref{fig02}. {Hence a
fit to the low concentration limit of the Langmuir model seems
reasonable. Unfortunately, one of the points deviating from }
Eq.~(\ref{langmuir_linear}) {is the PM intensity, which is used
as reference measure. In order to calibrate the fit correctly one
should reweight the reference PM intensity.  We therefore fit the
data against} Eq.~(\ref{how_to_fit}) {using instead of the actual
PM intensity as a reference, a rescaled value $I_{PM}^* = \alpha
I_{PM}$, which is the value the PM intensity would have if the data
would agree with} Eq.~(\ref{langmuir_linear}) {in the whole
intensity range. We estimate $\alpha$ from the crossing of the
$50$pM fitting line in} Fig.~\ref{fig02}(a) {with the $\Delta
\Delta G =0$ axis. This estimate is $\alpha=30$.  The effects of a
change in $\alpha$ on the fitting parameters will be discussed
below.}

Figures \ref{fig04}(a-d) show the result of the fit to
Eq.~(\ref{how_to_fit}), {using $\alpha=30$. In the main frames
each experiment is fitted independently.  In the insets the free
energy parameters are obtained from a simultaneous fit of all $50$pM
experiments. The latter data produce more accurate parameters, as
they come from using $4$ independent experiments (the 4 experiments
at $50$ pM, oligo synthesis a, in Table 3), hence the $58$
parameters are obtained on sampling over $1006 \times 4$ data
points. Both the free energy range and the continuity of the curves
in} Fig.~\ref{fig04} {are now as expected. The data show very
little spreading in comparison with the curves in} Figure
\ref{fig02}(a). {A quantification of the spreading for a
monotonic curve can be assessed by the Spearman's rank correlation
coefficient, which for all four experiments is very close to $1$.
This is an indicator of the reliability of the the nearest neighbor
fitted parameters. The ratio of data points over tuning parameters
is large 4024/58, which ought to yield a reliable fit. Moreover,
although the data are fitted to a linear model, all four experiments
show a clear deviation for the highest intensities. This is an
indication against overfitting, which would result in a fully linear
curve with erroneous fitting parameters. Therefore we conclude that
the deviations from the Langmuir isotherm observed in all four
experiments is a robust feature of the system and that the resulting
free energy parameters are physically meaningful. We also verified
that the free energy parameters obtained from the fit are quite
stable whether one fits the whole set of experimental data, or
whether the fit is restricted to the lowest intensity scales (e.g.
$I/I_{PM}^* \leq 5 \cdot 10^{-3}$) where all data clearly follow}
Eq.~(\ref{langmuir_linear}). {This is because the large majority
of experimental points in} Fig.~\ref{fig04} {are located in the
lowest intensity scales, anyhow. Hence, this additional data
filtering has little effects on the parameters.}

\begin{sidewaystable}[h]
{\scriptsize \caption{Free energy differences $\Delta \Delta G$
unique parameters obtained from fitting microarray data to
Eq.~(\ref{how_to_fit}). The data refer to triplets with central
mismatching nucleotides and flanking matching nucleotides. The
convention is that the numbers correspond for say, a mismatch
${AGT}\atop{TTA}$ to a free energy difference $\Delta
G\left({{AGT}\atop{TTA}}\right) -
 \Delta G\left({{AAT}\atop{TTA}}\right)$.
The upper strand has orientation 5' - 3'. The error bar on the
parameters is 0.2.}
 \centering
 \begin{tabular}{||c|c|c c c c||c|c c c c||c|c c c c||c|c c c c||}

\hline
\hline
\backslashbox{X}{Y}&  & A & C & G & T &
         &  A & C & G & T &
         &  A & C & G & T &
         &  A & C & G & T \\
\hline
\hline
A & \multirow{4}{*}{${XAY}\atop{X'AY'}$} & 2.2 & 2.0 & 2.4 & 2.2
  & \multirow{4}{*}{${XAY}\atop{X'CY'}$} & 3.0 & 2.8 & 3.0 & 3.0
  & \multirow{4}{*}{${XAY}\atop{X'GY'}$} & 2.5 & 1.8 & 2.5 & 2.2
  & \multirow{4}{*}{${XCY}\atop{X'AY'}$} & 2.4 & 2.2 & 2.4 & 2.5\\
C &                                    & 2.3 & 2.1 & 2.5 & 2.4 &
                                       & 3.0 & 2.8 & 3.0 & 3.0 &
                                       & 2.5 & 1.7 & 2.5 & 2.1 &
                                       & 2.4 & 2.2 & 2.4 & 2.5  \\
G &                                    & 1.9 & 1.8 & 2.2 & 2.0 &
                                       & 2.7 & 2.5 & 2.7 & 2.7 &
                                       & 2.4 & 1.6 & 2.4 & 2.0 &
                                       & 2.0 & 1.8 & 2.1 & 2.1 \\
T &                                    & 2.2 & 2.1 & 2.5 & 2.3 &
                                       & 3.1 & 2.9 & 3.1 & 3.1 &
                                       & 2.4 & 1.7 & 2.5 & 2.1 &
                                       & 2.4 & 2.2 & 2.4 & 2.5 \\
 \hline
 \hline
A & \multirow{4}{*}{${XCY}\atop{X'CY'}$} & 3.9 & 3.4 & 3.4 & 4.0
  & \multirow{4}{*}{${XCY}\atop{X'TY'}$} & 2.5 & 2.4 & 2.4 & 2.8
  & \multirow{4}{*}{${XGY}\atop{X'AY'}$} & 1.5 & 1.3 & 1.7 & 1.7
  & \multirow{4}{*}{${XGY}\atop{X'GY'}$} & 2.4 & 1.8 & 2.3 & 1.9\\
C &                                    & 3.4 & 3.0 & 2.9 & 3.5 &
                                       & 2.4 & 2.3 & 2.3 & 2.7 &
                                       & 1.7 & 1.6 & 1.9 & 1.9 &
                                       & 2.7 & 2.1 & 2.6 & 2.2  \\
G &                                    & 3.1 & 2.7 & 2.7 & 3.2 &
                                       & 2.5 & 2.5 & 2.5 & 2.8 &
                                       & 1.1 & 0.9 & 1.3 & 1.3 &
                                       & 2.5 & 1.9 & 2.4 & 2.0  \\
T &                                    & 3.8 & 3.4 & 3.3 & 3.9 &
                                       & 2.5 & 2.5 & 2.4 & 2.8 &
                                       & 1.7 & 1.6 & 2.0 & 2.0 &
                                       & 2.8 & 2.2 & 2.7 & 2.3 \\
 \hline
 \hline
A & \multirow{4}{*}{${XGY}\atop{X'TY'}$} & 2.0 & 1.8 & 1.9 & 1.9
  & \multirow{4}{*}{${XTY}\atop{X'CY'}$} & 3.5 & 3.6 & 3.1 & 3.2
  & \multirow{4}{*}{${XTY}\atop{X'GY'}$} & 2.2 & 2.2 & 2.0 & 2.4
  & \multirow{4}{*}{${XTY}\atop{X'TY'}$} & 2.3 & 2.4 & 2.0 & 2.2 \\
C &                                    & 1.6 & 1.4 & 1.5 & 1.5 &
                                       & 3.2 & 3.3 & 2.8 & 3.0 &
                                       & 2.3 & 2.3 & 2.1 & 2.5 &
                                       & 2.1 & 2.2 & 1.7 & 2.0 \\
G &                                    & 1.8 & 1.7 & 1.8 & 1.7 &
                                       & 3.1 & 3.2 & 2.8 & 2.9 &
                                       & 2.4 & 2.4 & 2.2 & 2.6 &
                                       & 2.4 & 2.5 & 2.1 & 2.4\\
T &                                    & 1.6 & 1.4 & 1.6 & 1.5&
                                       & 3.2 & 3.3 & 2.9 & 3.0 &
                                       & 2.3 & 2.3 & 2.1 & 2.5 &
                                       & 2.2 & 2.3 & 1.9 & 2.1\\
 \hline
 \hline

 \end{tabular}
 \label{DDG_marray}
}
 \end{sidewaystable}

\begin{sidewaystable}[h]
{\scriptsize \caption{Data as in Table~\ref{DDG_marray} using the nearest
neighbor parameters obtained from melting experiments in solution (see
\cite{sant04} and references therein). The data are at $T=65^\circ$ C
and at $1$ M [Na$^+$].}
 \centering
 \begin{tabular}{||c|c|c c c c||c|c c c c||c|c c c c||c|c c c c||}

\hline \hline \backslashbox{X}{Y}&  & A & C & G & T &
         &  A & C & G & T &
         &  A & C & G & T &
         &  A & C & G & T \\
\hline \hline
A & \multirow{4}{*}{${XAY}\atop{X'AY'}$} & 1.3 & 2.0 & 2.3 & 2.0
  & \multirow{4}{*}{${XAY}\atop{X'CY'}$} & 2.9 & 3.6 & 3.5 & 2.6
  & \multirow{4}{*}{${XAY}\atop{X'GY'}$} & 2.3 & 1.7 & 2.9 & 1.8
  & \multirow{4}{*}{${XCY}\atop{X'AY'}$} & 1.4 & 1.8 & 2.1 & 1.8 \\
C &                                    & 1.6 & 2.3 & 2.6 & 2.2 &
                                       & 3.5 & 4.2 & 4.1 & 3.2 &
                                       & 2.7 & 2.0 & 3.2 & 2.1 &
                                       & 2.2 & 2.6 & 3.0 & 2.6 \\
G &                                    & 1.6 & 2.3 & 2.6 & 2.3 &
                                       & 3.1 & 3.8 & 3.7 & 2.9 &
                                       & 2.6 & 2.0 & 3.2 & 2.1 &
                                       & 2.1 & 2.5 & 2.9 & 2.6 \\
T &                                    & 1.1 & 1.8 & 2.1 & 1.8 &
                                       & 3.0 & 3.7 & 3.6 & 2.7 &
                                       & 2.5 & 1.8 & 3.0 & 1.9 &
                                       & 1.9 & 2.3 & 2.6 & 2.3 \\
 \hline
 \hline
A & \multirow{4}{*}{${XCY}\atop{X'CY'}$} & 3.4 & 4.3 & 4.4 & 4.5
  & \multirow{4}{*}{${XCY}\atop{X'TY'}$} & 2.2 & 2.5 & 2.3 & 2.2
  & \multirow{4}{*}{${XGY}\atop{X'AY'}$} & 0.8 & 0.8 & 1.3 & 1.0
  & \multirow{4}{*}{${XGY}\atop{X'GY'}$} & 2.1 & 1.6 & 2.6 & 1.7 \\
C &                                    & 3.6 & 4.5 & 4.5 & 4.7 &
                                       & 2.7 & 3.0 & 2.8 & 2.7 &
                                       & 1.6 & 1.6 & 2.1 & 1.7 &
                                       & 2.8 & 2.2 & 3.2 & 2.4 \\
G &                                    & 3.1 & 4.0 & 4.1 & 4.2 &
                                       & 2.3 & 2.6 & 2.4 & 2.4 &
                                       & 0.7 & 0.7 & 1.1 & 0.8 &
                                       & 2.1 & 1.5 & 2.6 & 1.7 \\
T &                                    & 2.6 & 3.5 & 3.6 & 3.7 &
                                       & 2.1 & 2.4 & 2.2 & 2.2 &
                                       & 1.4 & 1.3 & 1.8 & 1.5 &
                                       & 2.3 & 1.7 & 2.8 & 1.9 \\
 \hline
 \hline
A & \multirow{4}{*}{${XGY}\atop{X'TY'}$} &  2.0 & 1.7 & 1.7 & 1.7
  & \multirow{4}{*}{${XTY}\atop{X'CY'}$} & 3.3 & 3.6 & 3.6 & 3.1
  & \multirow{4}{*}{${XTY}\atop{X'GY'}$} & 2.4 & 2.2 & 2.4 & 2.5
  & \multirow{4}{*}{${XTY}\atop{X'TY'}$} & 2.5 & 2.8 & 2.4 & 2.6 \\
C &                                    & 1.6 & 1.3 & 1.3 & 1.3 &
                                       & 3.6 & 4.0 & 4.0 & 3.4 &
                                       & 2.5 & 2.2 & 2.4 & 2.6 &
                                       & 2.3 & 2.6 & 2.2 & 2.4 \\
G &                                    & 1.4 & 1.1 & 1.1 & 1.1 &
                                       & 3.6 & 3.9 & 3.9 & 3.4 &
                                       & 2.6 & 2.4 & 2.6 & 2.7 &
                                       & 2.4 & 2.8 & 2.4 & 2.6 \\
T &                                    & 1.4 & 1.1 & 1.1 & 1.1 &
                                       & 3.2 & 3.6 & 3.6 & 3.0 &
                                       & 2.5 & 2.2 & 2.4 & 2.6 &
                                       & 1.8 & 2.2 & 1.7 & 2.0 \\
 \hline
 \hline

 \end{tabular}
 \label{DDG_santa}
}
 \end{sidewaystable}

{Table} \ref{DDG_marray} {shows the free energy parameters $\Delta
\Delta G_{\rm \mu array}$ as obtained from the above fitting procedure.
Because of the degeneracies mentioned above (see e.g.} Eq.~(\ref{reparam})
and Ref.~\cite{gray97a}) {the dinucleotide parameters are not uniquely
determined. Triplet parameters are however unique, and these are given
in the Table. The parameters are the $\Delta \Delta G$ defined, for
instance, as:}

\begin{equation}
\Delta \Delta G
\left({{\rm A {\underline C} G}}
\atop {{\rm T {\underline T} C}}  \right)
=
\Delta G
\left({{\rm A {\underline C} G}}
\atop {{\rm T {\underline T} C}}  \right)
- \Delta G
\left({{\rm A A G}}
\atop {{\rm T T C}}  \right)
\end{equation}
{where the upper strand is 5'-3' oriented. The lower strand is
the invariant target sequence, the upper strand are the probe
sequences. Hence the $\Delta \Delta G$ parameters are measured
subtracting the reference perfect match probe. Note that because of
this subtraction one has}
\begin{equation}
\Delta \Delta G \left({{\rm A {\underline C} G}} \atop {{\rm T
{\underline T} C}}  \right) \neq \Delta \Delta G \left({{\rm C
{\underline T} T}} \atop {{\rm G {\underline C} A}}  \right)
\end{equation}
{as the reference PM sequence is different in the two cases.

Using standard linear regression tools we estimated the error bar on
the parameters of Table} \ref{DDG_marray} {to be equal to $0.2$.
In order to compare with existing published data} \cite{sant04} {we
present in Table} \ref{DDG_santa} the $\Delta \Delta G_{sol}$ {for
triplets following the same notation as in Table} \ref{DDG_marray}. {As
mentioned before the data in solution are at $T=65^\circ C$ and 1M
[Na$^+$].} Figure~\ref{fig05} {shows a plot of the two free energies
$\Delta \Delta G_{\rm \mu array}$ vs.  $\Delta \Delta G_{\rm sol}$. A
clear quantitative correlation between the two is observed. The Pearson
correlation coefficient is $0.839$. In comparing the two sets we note
that the 16 mismatches of CC appear to be the most deviating in the
two cases.}

{As discussed above, the fit was done with a rescaled PM
intensity, using a factor $\alpha=30$. We have repeated the analysis
for other values of $\alpha$. Varying $\alpha$ causes a global shift
of the data in Table} \ref{DDG_marray} {by an $\alpha$ dependent
constant. This shift does not affect the slope or correlation of the
data in} Fig.~\ref{fig05}. By using $\alpha=50$ we found a positive
shift of $0.17$, while setting $\alpha=20$ produces a shift of
$-0.14$. These two values of $\alpha$ are our estimate of the
largest range of variability for this parameters. In general the
procedure of reweighting the PM intensity with $\alpha$ introduces a
global error $\pm 0.2$ affecting all parameters in Table
\ref{DDG_marray}.

\subsection*{Concluding remarks}

During the past decades a considerable amount of research was devoted
to the quantification of interactions among hybridizing nucleic acid
strands in aqueous solution. This lead to a parametrization, via the
nearest-neighbor model, of the contribution to the total free energy in
terms of dinucleotide pairs for perfect matching DNA/DNA \cite{sant98},
RNA/RNA \cite{xia98} and DNA/RNA \cite{sugi95} duplexes, but also for
strands with an internal mismatch \cite{sant04}.  This large amount
of data is currently used in various applications as for instance for
calculation of DNA melting temperatures or for RNA secondary structure
predictions.  As it has been widely recognized \cite{sant04,fish07,pozh07}
a similar effort for quantifying interactions in DNA microarrays is very
important. This effort will lead to a better understanding of molecular
interactions in DNA microarrays and ultimately on their functioning.

A precise quantification of interactions bring{s} some
challenges. First of all many different microarray platforms exist,
they differ by the length of probe sequences and the way these are
covalently linked to the solid surface. It is not unlikely that
interactions between hybridizing strands { are} of slightly
different nature in these different platforms.  Hence, one should be
careful for instance to generalize the results of this work to, say,
Affymetrix GeneChips. In addition, in order to measure accurately
interaction parameters, one needs a careful experimental setup in
which possible competing reactions, as hybridization between
partially complementary strands in solution, are absent. In the case
of the present work this was achieved by choosing a single sequence
in solution hybridizing with perfect matching probe sequences with
one or two internal mismatches. It is difficult to directly fit the
free energy parameters from complex biological experiments where the
hybridizing solution contain a large number of interacting
sequences. This may explain why in some cases poor correlations
between $\Delta G_{\rm sol}$ and $\Delta G_{\rm \mu array}$ was
reported \cite{naef03,zhan03}. { One of the advantages of the
experimental setup chosen in this work is that one can obtain in
principle all parameters in a single experiment, as all
hybridization reactions with one or two mismatches occur in
``parallel" on a single array. However, a drawback is that in this
setup one can determine only the free energy and not the
contribution of enthalpy and entropy separately, which would allow
to extend the parameters to other temperatures.  It would be
certainly interesting to extend the analysis to other platforms and
hybridization conditions.}

{ In the present work we focused on the determination of $\Delta
\Delta G$ which is the free energy difference between a perfect
matching hybridization and an hybridization where the probe
sequences has one or more internal mismatches. Quantifying the
effect of internal mismatches is important for a better
understanding of cross-hybridization effect, which is the unintended
binding of non-perfectly complementary sequences to a given probe.
Moreover, this understanding could have some practical consequences
for optimal probe design. An advantage of the parameter $\Delta
\Delta G$ is that it is insensitive to the free energy initiation
parameter} (Eq. (\ref{DGpm})) {and the scaling factor $A$} (Eq.
(\ref{langmuir}), Eq. (\ref{langmuir_linear})) {and that it is
expected to be less sensitive to buffer conditions as ionic salt etc
\ldots The determination of the perfect match parameters $\Delta G$ is
also possible in principle from microarray experiment but it requires
sampling perfect match hybridizations from a large number of target
sequences. This requires a different and more complex experimental design.}

\begin{figure}[ht]
\begin{center}
\rotatebox{0}
{\scalebox{0.35}
{\includegraphics*{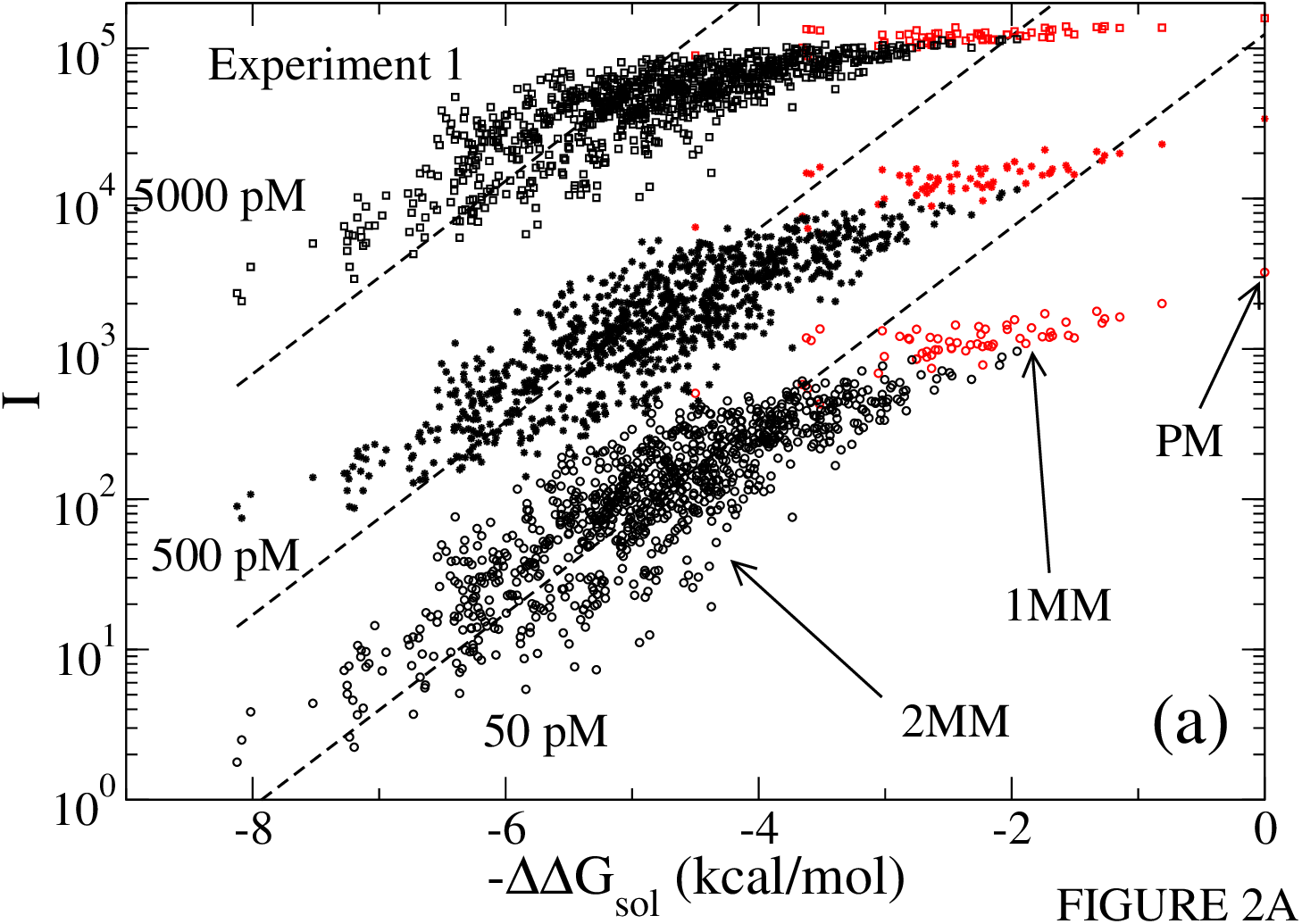}
\includegraphics*{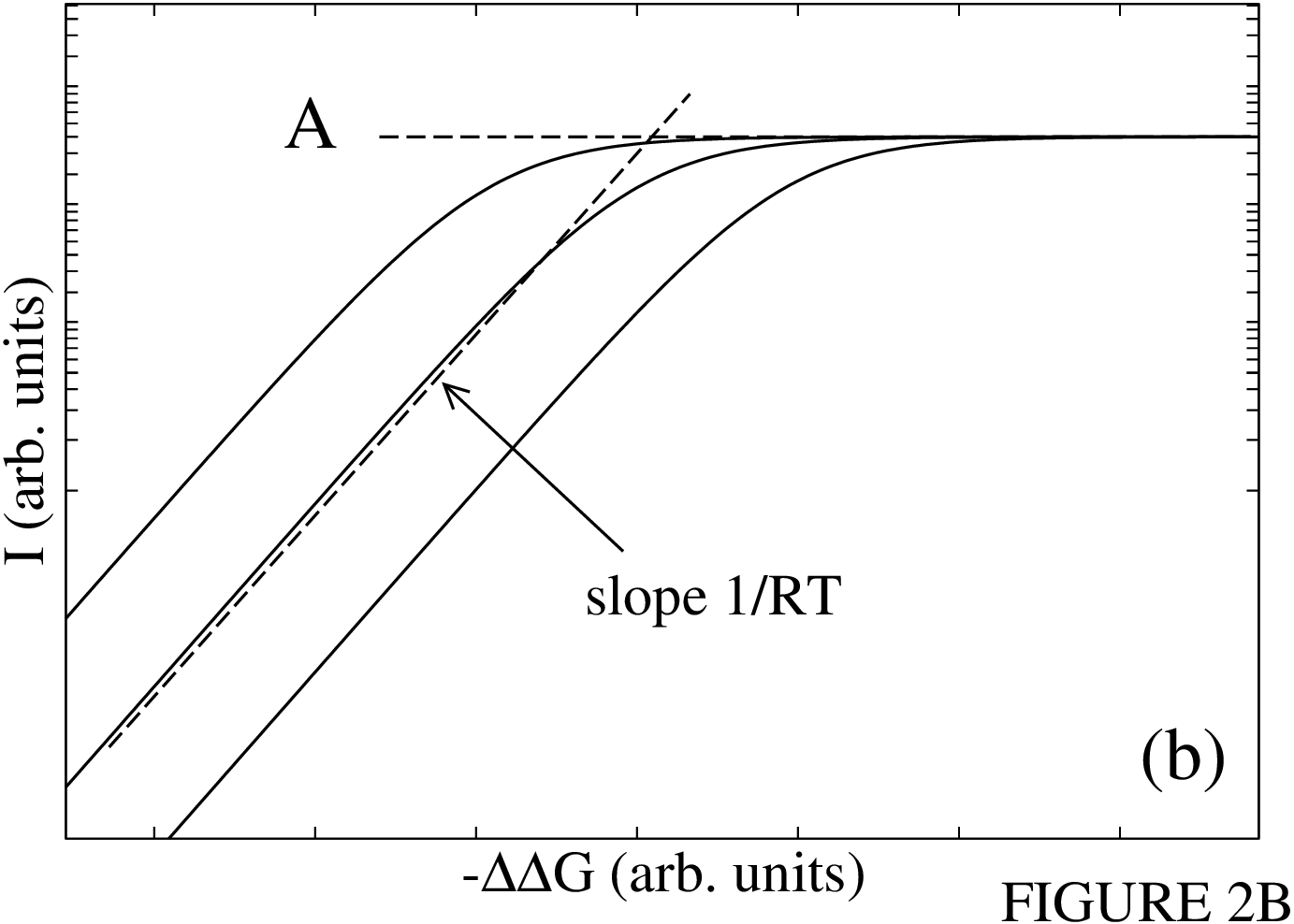}}
}

%
\end{center}
\caption{(a) Plot of the intensities as functions of $\Delta \Delta
G_{\rm sol}$, the difference of hybridization free energy with
respect to the perfect match free energies, from nearest neighbor
free energies obtained from melting experiments in solution.  With
this choice of parameters the perfect match is {\bf located at
$\Delta \Delta G =0$.}
The different plots correspond
to concentrations of $50$, $500$ and $5000$ pM (from bottom to top).
The lines drawn have slopes corresponding to $1/RT$, with $T=65^\circ C =
338 K$ the experimental temperature.  (b) Behavior of three concentration
data as predicted from the Langmuir model (Eq.~(\ref{langmuir})).
}
\label{fig02}
\end{figure}

\begin{figure}[ht]
\begin{center}
\rotatebox{0} {\scalebox{0.45} {\includegraphics*{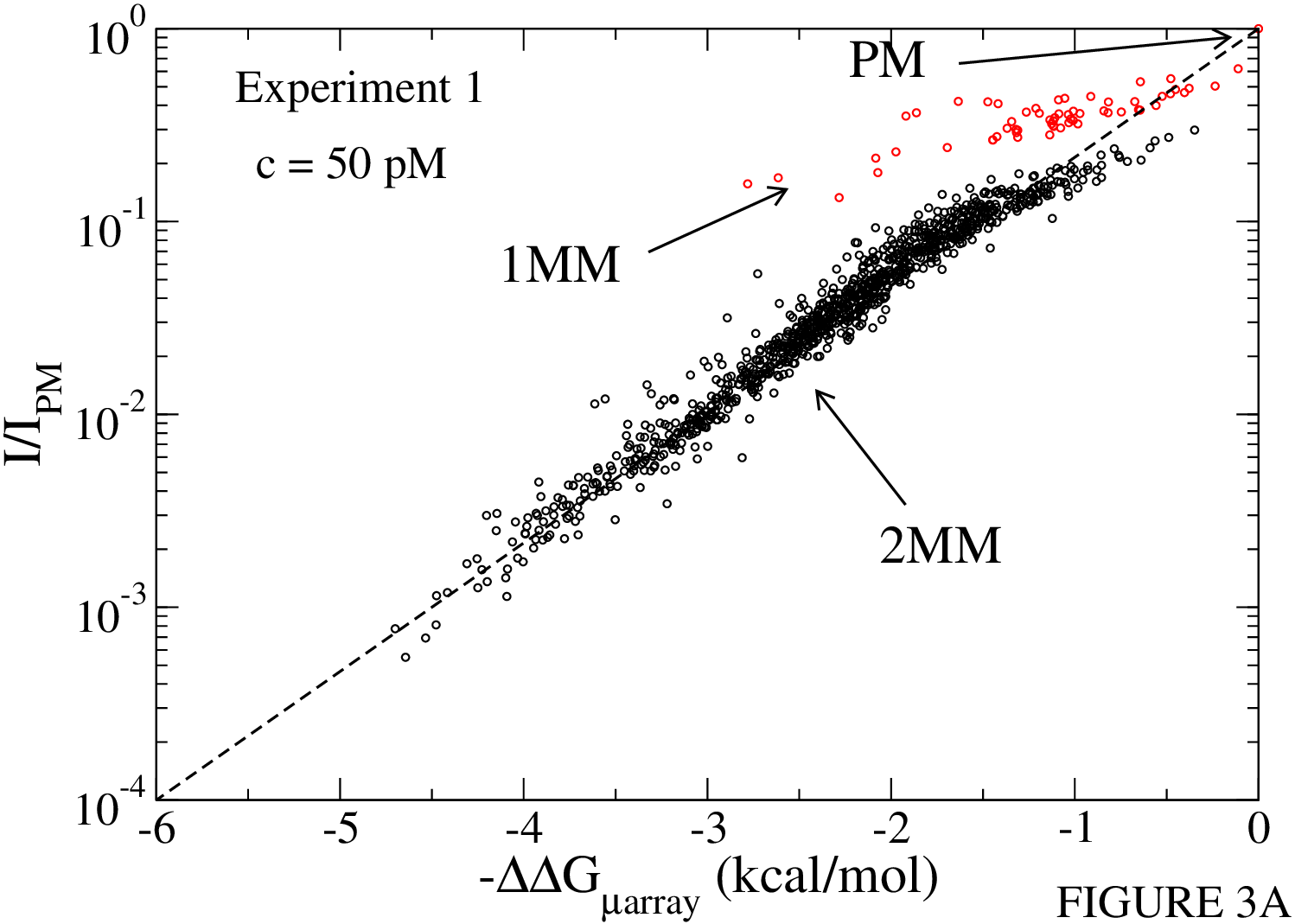} }}

\end{center}
\caption{Ratios of Intensities and perfect match intensities vs.
$\Delta \Delta G_{\rm \mu array}$, the {\bf relative} hybridization
free energy between two strands as obtained from a fit to
Eq.~(\ref{how_to_fit}). Three distinct groups of points are
indicated: PM for perfect match, 1MM for probes with a single
internal mismatch and 2MM for probes with two mismatches. The dashed
line in is drawn as a guide to the eye.} \label{fig03}
\end{figure}

The present work on custom Agilent arrays shows that there is a
strong correlation, also on the quantitative scale, between $\Delta
\Delta G_{\rm sol}$ and $\Delta \Delta G_{\rm \mu array}$.
This correlation is
{ shown in} Fig.~\ref{fig05} {with explicit free energy values
given in Tables} \ref{DDG_marray} and \ref{DDG_santa}.
 A fit of the interaction parameters from microarray data
shows a much better agreement of the data with the thermodynamic
models (compare Fig.~\ref{fig02} with Fig.~\ref{fig03}). However, in
absence of dedicated experiments for the determination of
interaction free energies { on a DNA microarray}, the results of
this work suggest that one could use as
{approximations for them the corresponding} hybridization free
energies in solution. {Recent work} \cite{weck07,fish07} {has
addressed the issue of the correlation between $\Delta G_{\rm sol}$
and $\Delta G_{\rm \mu array}$. Ref.} \cite{weck07} {considered
oligonucleotide microarrays on Codelink activated slides carrying
one, two or three mismatches. The data plotted as a function of
$\Delta G_{\rm sol}$ showed a good agreement with the Langmuir
model, implying a fair correlation between $\Delta G_{\rm sol}$ and
$\Delta G_{\rm \mu array}$. However the number of data points was
insufficient to perform a direct fit of the thermodynamic parameters
from the microarray data. Interestingly, the lowest concentration
data in Ref.} \cite{weck07} {seem to indicate the existence of
deviations from the Langmuir model similar to those observed in}
Fig.~\ref{fig02}(a). Fish et al. \cite{fish07} {performed a
series of experiments on oligo sequences in solutions hybridizing to
perfect match and to sequence carrying one to multiple mismatches.
Their analysis included tandem mismatches, i.e. mismatches on
neighboring sequence sites (in our case the minimal distance between
mismatches is five nucleotides).  An overall correlation between
$\Delta G_{\rm sol}$ and microarray intensities was observed,
implying a correlation between $\Delta G_{\rm sol}$ and $\Delta
G_{\rm \mu array}$.  In these experiments $\Delta G_{\rm sol}$ was
measured directly from experiments in solution and did not rely on
the nearest-neighbor model parameters.  As a correlation between
$\Delta G_{\rm sol}$ and $\Delta G_{\rm \mu array}$ has by now been
observed in several different microarray platforms, it is fair to
expect that such a correlation is a general feature of microarrays.
However, an accurate determination of nearest-neighbor parameters in
other platforms would be very useful for a better quantification of
this correlation.  }


{An interesting issue is the deviation from the low concentration limit
of the Langmuir model} (Eq.~(\ref{langmuir_linear})). {These deviations
cannot be explained by the full model of} Eq.~(\ref{langmuir}). {There
are several underlying approximations in the Langmuir model, as for
instance hybridization is always considered two state.  The model
also assumes that hybridizing strands, apart from forming a duplex,
do not further interact with other strands at the surface. Moreover,}
Eqs.~(\ref{langmuir}) and (\ref{langmuir_linear}) {apply to a system
in thermal equilibrium.  More investigations are necessary for a better
understanding on the deviation from the Langmuir model found in this
study. These will involve further experiments in different external
conditions, e.g. different temperatures or salt concentrations as well
as theoretical analysis, which are left for some future work.}

\begin{figure}[ht]
\begin{center}
\rotatebox{0}
{\scalebox{0.35}
{\includegraphics*{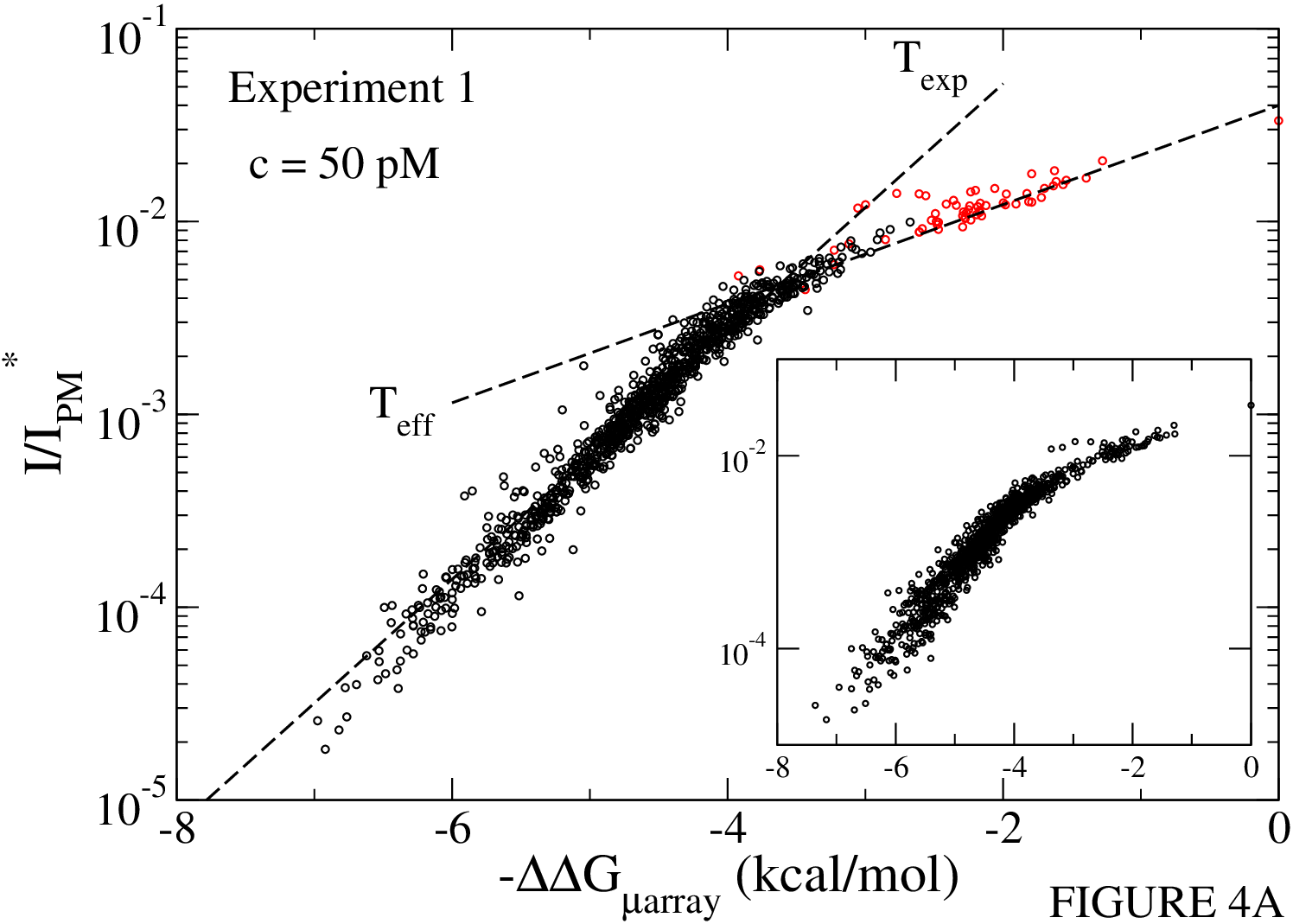}
 \includegraphics*{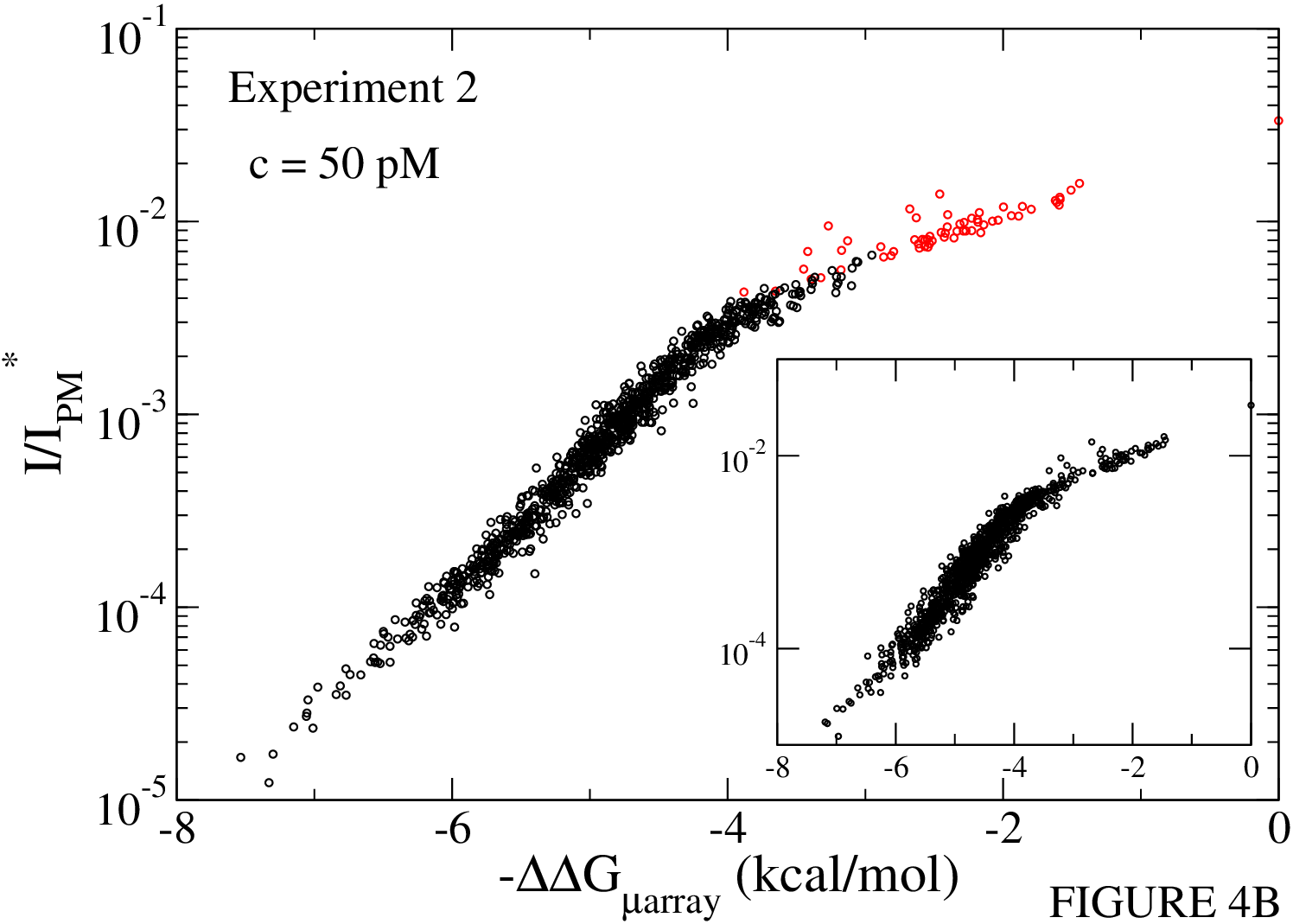}}}

{\scalebox{0.35}
{\includegraphics*{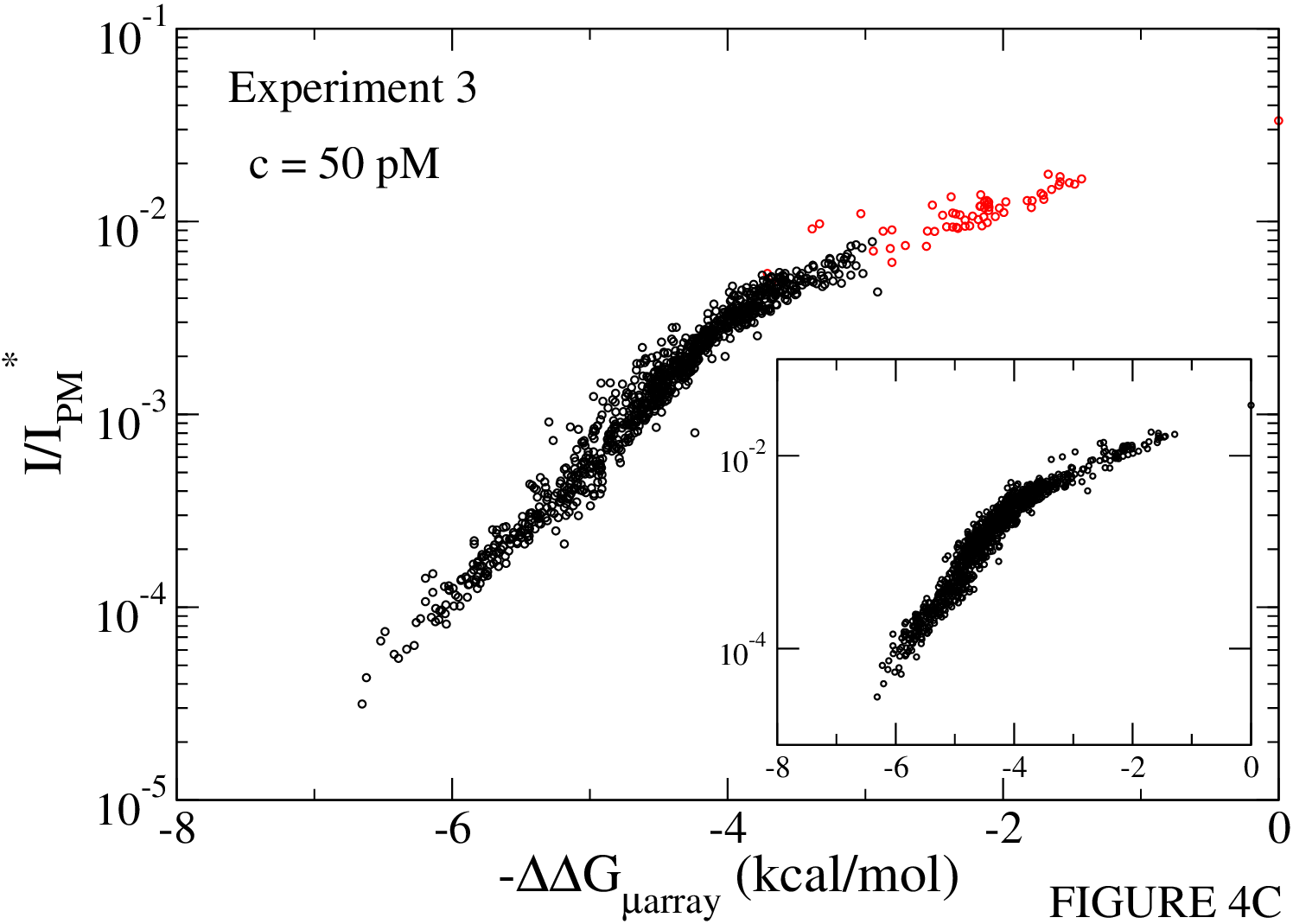}
 \includegraphics*{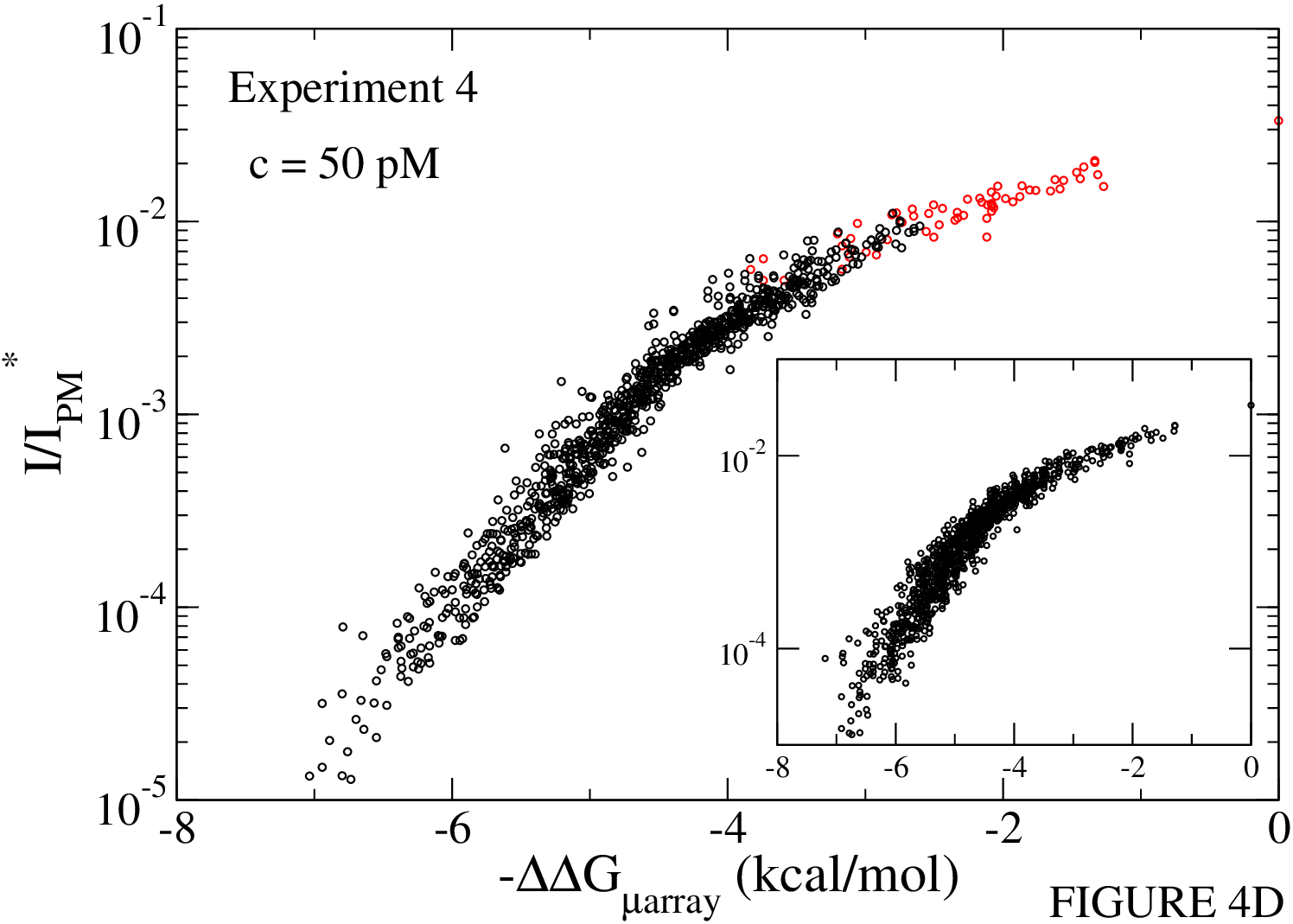}}}
\end{center}
\caption{{Plot of $I/I_{\rm PM}^*$ where $I_{\rm PM}^* = \alpha
I_{\rm PM}$ (where we took $\alpha=30$ as explained in the text) as
function of the nearest neighbor fitted $\Delta \Delta G_{\rm \mu
marray}$. The alignment of the intensities onto single monotonic
curves is a proof of the good quality of the fits.  In the main
frame the four different experiments where fitted separately. The
insets show the date from intensities of each experiments, but the
fit was done globally on all experiments at $50$ pM. As a
measurement of the goodness of the fits the Spearman's rank
correlation coefficient was used. This coefficient is for the main
frames plots (a-d): 0.9860 0.9911 0.9866 and 0.9867. For the four
plots in the insets the correlation coefficients are: 0.9732 0.9705
0.9748 and 0.9699.  The two straight lines in the first main frame
correspond to slopes $1/RT$ where we took $T_{exp} = 65^\circ \ {\rm
C} = 338 K$ for the experimental temperature and $T_{eff}=850$K.}}
\label{fig04}
\end{figure}

\begin{figure}[ht]
\begin{center}
\rotatebox{0}
{\scalebox{0.45}
{\includegraphics*{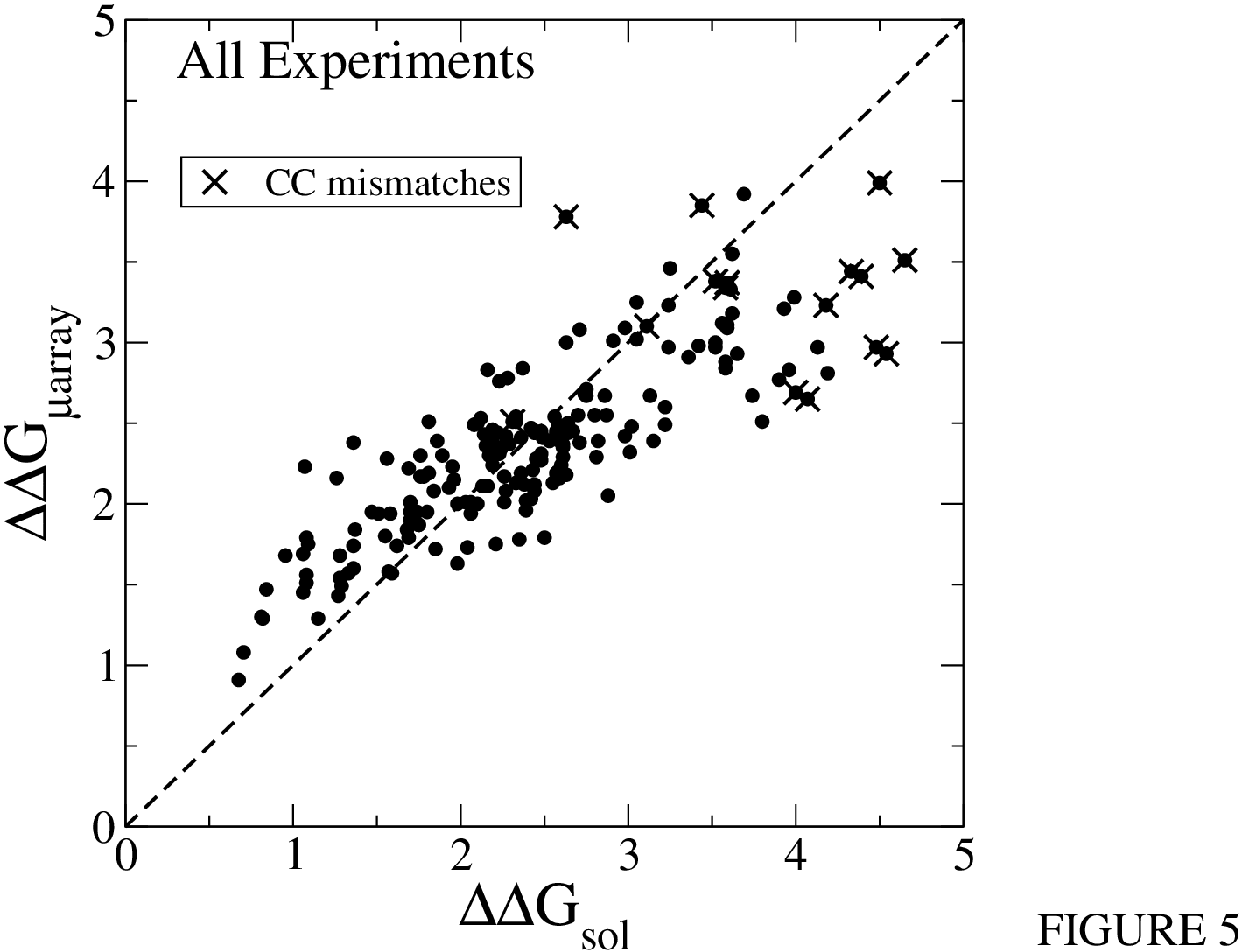}}}
%
\end{center}
\caption{Comparison of data in Table \ref{DDG_marray} and
\ref{DDG_santa}: the free energy differences between a perfect
matching hybridization and hybridization with an internal mismatch
as obtained from data from Ref.~\cite{sant04} ($\Delta \Delta G_{\rm
sol}$) and from a fit of the microarray data ($\Delta \Delta G_{\rm
\mu array}$). The results show a good quantitative correlation
between the two quantities:  the Pearson correlation coefficient is
0.839 } \label{fig05}
\end{figure}

{It is interesting to remark that the deviation from the Langmuir model
``enhances" the cross-hybridization problem because there is a smaller
effect on intensity for a given free energy penalty (smaller slope in}
Figure \ref{fig04}). {As an example, a mismatch with $\Delta \Delta
G = 2.5$ kcal/mol (a typical value from} Table~\ref{DDG_marray}) {
corresponds to a $I/I_{\rm PM}$ ratio of $\approx 0.02$ in the regime
governed by the Langmuir model, compared to $\approx 0.2$ in the
deviating regime. This implies that in the deviating regime a significant
fraction of the amount of target binding to a PM probe binds to a
probe carrying one internal mismatch.}

{Although the origin of these deviations are not known it is
remarkable that the data appear to follow approximately two straight
lines separated by a sharp kink} (Fig.~\ref{fig04}). {Although extensions
of the Langmuir model in the context of DNA microarrays have been
discussed} (see e.g. \cite{vain02}) {we are unaware of isotherms
which could have a shape as shown in} Fig.~\ref{fig04}.{
The presence of a second straigth line in the $\log $ plot implies
that in this range the data still follow the thermodynamic model of}
Eq.~(\ref{langmuir_linear}) {but with a different ``effective"
temperature than the experimental one. A linear regression to the data
yields $T_{\rm eff} \approx 850K$, which is higher than the experimental
temperature.  It is interesting to point out that recent analysis}
\cite{held03,carl06} {of Affymetrix GeneChip data use Langmuir
model with $\Delta G_{\rm sol}$ rescaled to higher effective higher
temperatures. A better understanding of the regime governed by an
effective temperature may provide new insights on this issue.}

\begin{ack}
We thank Kizi Coeck {and Karen Hollanders} for help with the
experiments. We aknowledge financial support from the Fonds voor
Wetenschappelijk Onderzoek (FWO) under grant G.03111.08.
\end{ack}


\raggedright 


\clearpage

\clearpage

\clearpage

\clearpage

\clearpage

\end{document}